\documentclass[12pt]{article}

\usepackage[english]{babel}									
\usepackage[utf8]{inputenc}		
\usepackage{tabularx}
\usepackage[T1]{fontenc}		
\usepackage{comment}
\usepackage{amsmath,amsthm,amsfonts,amssymb}
\usepackage{setspace} 
\usepackage{hyperref}

\title{\normalsize\bf%
\uppercase{Bitcoin: A new proof-of-work system with reduced variance}
}
\author{%
Danilo Bazzanella$^{1}$,   Andrea Gangemi$^{1}$
}

\begin{document}

\date{}

\maketitle

\vspace{-0.5cm}

\begin{center}
{\footnotesize 

$^1$Dipartimento di Scienze Matematiche, Politecnico di Torino,\\ Corso Duca degli Abruzzi 24, 10129 Torino — Italy\\
E-mails: danilo.bazzanella@polito.it / andrea.gangemi@polito.it  
}
\end{center}


\begin{abstract}
\textit{Proof-of-Work (PoW) is a popular consensus protocol used by Bitcoin since its inception. PoW has the well-known flaw of assigning all the reward to the single miner (or pool) that inserts the new block. This has the consequence of making the variance of the reward and thus the mining enterprise risk extremely high. To address this problem, Shi in 2016 proposed a theoretical algorithm that would substantially reduce the issue.  We introduce a variant of Proof-of-Work that improves on Shi's idea and can be easily implemented in practice. In order to insert a block, the network must not find a single nonce, but must find a few of them. This small change allows for a fairer distribution of rewards and at the same time has the effect of regularizing the insertion time of blocks. This would facilitate the emergence of small pools or autonomous miners.}
\end{abstract}


\baselineskip=\normalbaselineskip

\section{Introduction}

\textit{Proof of Work} (PoW) is a form of cryptographic proof in which one party (the prover) demonstrates  to the others (the verifiers) that a large amount of computational work has been done. An essential condition is that the verification must be done with little computational effort.

The basic idea was proposed by Cynthia Dwork and Moni Naor in 1992 \cite{Dwork-Naor92} to deter denial-of-service attacks and email spam, and consists of requiring the user of a service to perform a certain amount of computation, which is time-consuming to perform but quick to verify, before being able to access the service.

The term Proof of Work  was formalised a few years later, in a 1999 paper written by Markus Jakobsson and Ari Juels \cite{Jako-Jules}, but the most successful version of proof of work is due to Adam Back, who proposed in 2002 the protocol called \textit{Hashcash} in a technical report entitled \textit{A Denial of Service Counter-Measure} \cite{Back}.

Hashcash is very simple, powerful and flexible. The idea is to take any text, add a small piece of random text, the so-called \textit{nonce}, and require the hash of the composed text to be below a certain target. If it is not below the target, one must try again with a new nonce, until a hash below the target is obtained. Since it is impossible to predict the output of the hash function as the input text changes, the protocol works like randomly extracting a number in a certain range and requiring it to fall below a certain value, which makes it easy to check the probability  of success.
Verifying that a large amount of calculation has actually been done is straightforward: just compute a single hash with the nonce that is shown by the person who performed the PoW and verify that it is below the set target. 

Bitcoin's consensus protocol is based on the hashcash protocol and works as follows. Each miner builds a possible block of transactions and then tries to add it to the blockchain, searching a nonce to insert into the block so that the hashing of the block header is less than a public target, set by the network protocol. The miner who first obtains a valid nonce has the right to insert the new block and receive the reward (6.25BTC in 2022). Every 2016 blocks, roughly two weeks, the target is modified in order to keep the block generation time at 10 minutes on average, regardless of the total computing power of the network \cite{Antonopoulos}.

This consensus protocol guarantees high security. A user would have to have more computing power than the rest of the network to be able to attack the network, the so-called \textit{51\% attack}. The drawback is that it is very complicated to add a block to the blockchain without a major investment in adequate equipment, such as Application Specific Integrated Circuits (\textit{ASICs}). This greatly discourages autonomous mining and pushes all users to join a mining pool, paying fees to reduce the business risk.

One problem with this natural tendency is that, currently, the 6 largest pools together account for more than $80\%$ of the computing power of the entire network \cite{pools}, which makes the network very poorly decentralised.

Another problem stems from the fact that, in  Bitcoin, the miner (or pool) that inserts the block takes the whole reward, while all the other miners take nothing. 
This makes the variance in miners' profit very high and the only way to remedy to this effect is to be a huge miner, with the associated
investments, or to work for a pool.

Finally, in Bitcoin the insertion time of a block has a high variance. It is easy to verify, both mathematically and in practice, that only 7\% of blocks are produced between 9 and 11 minutes, while only 14\% of blocks are created between 8 and 12 minutes. It has been shown that the high variance of block insertion time is at the root of critical attacks on PoW blockchains \cite{Bobtail}.

It would be very important for the security of the network to modify the Bitcoin consensus protocol to promote autonomous mining even by not huge mining farms, while reducing the concentration of mining pools and the variance of blocking times. 

The aim of this paper is to propose flexible proof-of-work system that can provide better decentralisation of the network, redistributing rewards more fairly and lowering the variance of the time to insert a new block. The paper is organised as follows: Section 2 briefly describes  the previous work done by other researchers , Section 3 explains theoretically the idea for a new consensus protocol, and finally Section 4 reports a mathematical analysis of the new model.

\section{Related Work}

The idea of a more equitable distribution of the reward
for successfully inserting a block is not new. Sarkar
\cite{Sarkar} introduced in 2019 the \textit{Multi-stage Proof-of-Works}. The proof-of-work for mining a block is divided into multiple stages, and this has the effect to split the block reward into an equal number of stage rewards. Once a block gets onto the blockchain, the miner which successfully completed a particular stage can claim the reward for that stage. These types of protocols have been analysed in more detail in  \cite{Arco}. They proved that the mining probability might not be strictly related to the miner hashing power, and this could open up potential fairness issues in mining. In addition, some attacks have been proposed which appear to increase the vulnerability of the system.

A lot of work has also been done to discourage the creation of large Bitcoin mining pools. A line of research is trying to disincentivize large pools making it difficult to delegate the Proof-of-Work. Eyal and Sirer \cite{Eyal-Sirer} proposed the \textit{Two-Phase PoW}, that consists of two separate cryptopuzzles. The first one is identical to the one already existing on Bitcoin, so the miner must find a nonce such that the hash of the block header is less than a set target. In the second part, the block header must be signed with the coinbase transaction's private key, and the hash of this digital signature must be below a second difficulty parameter. Miller, Kosba, Katz and Shi, \cite{Miller} proposed  to disable mining pool enforcement mechanisms in a cryptographically strong manner, through \textit{strongly nonoutsourceable puzzles}.

A very interesting proposal, even if only theoretical, is the one proposed by Shi in \cite{Shi}. The idea is to find a problem with several solutions, rather than a single solution. The first miner (or pool) that finds all the solutions will cause the block to be inserted. The block will be inserted by one of the miners at random among those who have found at least one solution.

This is similar to the line taken by the Bobtail algorithm, suggested by Bissias  and Levine in 2020 \cite{Bobtail}. Bobtail is a possible concrete realization of Shi's model, but it is rather complex and difficult to manage. They proposed to maintain a public ranking of the lowest $k$ hashes in the network. The block is inserted by whoever is at the top of the ranking (i.e. has found the lowest hash), when the average of these $k$ values is less than a certain target $t_k$. If $k=1$, this is equivalent to the current Bitcoin, and the variance of interblock times decreases as $k$ increases. One of the advantages of this idea is that, by giving the ability to insert a block only to a specific miner, they have a number of forks comparable to that of the current Bitcoin protocol, even if $k > 1$. They also have an interesting section in which they describe possible new attacks against such generalised protocols.

We will see in the next section how our model is in some ways similar to Bobtail, but it is simpler and the number of communications is smaller.
Due to the greater simplicity of our new protocol, it is possible to mathematically prove that the expected value of the miners' profit remains unchanged compared to Bitcoin and its variance decreases.

\section{New Proof-of-Work systems}

Currently, the right to insert a new block into the Bitcoin blockchain and thus receive the reward is given to the miner (or pool) that first finds a nonce that, when inserted into the block, returns a hash of the block header that is less than a certain target. Our idea is to modify Bitcoin's proof-of-work protocol by searching for multiple nonces, each of which, when inserted into the producer block, returns a hash of the header that is less than the target. Each miner will have the opportunity to find multiple nonces and thus be rewarded in proportion to the number of nonces found. 

More precisely, one can set an integer parameter $j \geq 1$ and establish that a new block $B$ is inserted when the entire network finds $j$ nonces. Note that the case $j = 1$ is essentially the current Bitcoin protocol, but with a small difference: in fact, since it is in general necessary to reward multiple miners, it is not possible to do so with the coinbase transaction of the block $B$, since while proceeding with the PoW it is not yet known which miners will deserve a fraction of the block reward. Every nonce found must be published and verified by the network, and they must be inextricably linked to the Bitcoin address of the miner that found them. The coinbase transaction of the block $B + 1$ will then redistribute the rewards among the miners who found the nonces during the creation of the block $B$.

To maintain the desired block speed, we switch the current Bitcoin game to a repeated game that is $j$ times easier, but the block is inserted when the network wins $j$ times.

To link a nonce with the address of the miner who found it, we can use a hash function $h$, e.g. SHA256 already used by Bitcoin. The miners will look for a particular value $\text{nonce}_1$ to calculate another value, $\text{nonce}_2$, with the following formula:
\[\text{nonce}_2 = h(\text{miner address} \,\,||\,\, \text{nonce}_1).\]
$\text{nonce}_2$ is  the value that, when inserted into the block header, will return a digest that is below the set target.

When a miner finds a suitable nonce$_2$, he will publish his block, the value nonce$_1$ and his Bitcoin address.
The network is then able to  verify  that the PoW has been correctly executed.

Since every hash function is collision resistant, we can be reasonably sure that another miner will not be able to find another $\text{nonce}_1$ which, concatenated with his address,  will return the same value $\text{nonce}_2$.

Our model can be summarised as follows:

\begin{enumerate}
\item When a miner finds a correct nonce$_2$, he publishes the value nonce$_1$, the block header and the Bitcoin address where he will receive the reward. The network can easily verify the correctness of the published nonces;  
\item When the number of nonces published by the entire network equals j, the nonces are sorted and the miner who found the smallest nonce$_2$ gets the right to insert the new block. To prove possession of the address, the miner must use his digital signature. Rewards will be distributed in the next block;
\item The network starts to work for the insertion of a new block. The miners can insert the transactions they prefer, but the coinbase transaction will be fixed by the protocol and every miner is required to use the same. This shared coinbase distributes the reward plus the fees in proportion to the number of nonces found in the mining of the last inserted block;
\item The process starts again from the step 1).
\end{enumerate}

The easiest way to switch from the current Bitcoin protocol to our new protocol is to establish that a single block is entered without the coinbase transaction, postponing the reward distribution to the next block.

In our protocol we select a single miner to be the only one able to insert a block and this opens the way for the Denial of Service attack, already introduced in Bobtail \cite{Bobtail}. There is a simple solution to this problem. The network continues mining until a new block is inserted. If the miner with the smallest nonce$_2$ does not proceed to insert the block, a few moments later the network will find a new nonce and then there will be another miner able to insert the block (the one with the second smallest nonce$_2$). Any malicious miner who attempts to stop the system is therefore likely to lose his reward and in any case would not be able to stop the system for long.

For each choice of $j$, we have a system with the same average block insertion rate as the current protocol, but the variance of the insertion time is smaller and decreases as $j$ increases. It is also more likely that a block will be inserted in a time close to the average. See Section 4 for proofs about these facts.

The second advantage of these modified PoW models is that miners who do not insert a block, but who can still show that they worked for the system, are rewarded accordingly. The reward for each block is divided among the miners in proportion to the number of nonces they have found. 

A possible drawback of this new model are the forks. 
If two miners find the $j$-th nonce almost simultaneously, there are two users (the ones with the two smallest nonces) that are able to propose the insertion of their block to the network. The network is then left with two different last blocks (a fork). Each miner then resumes mining by choosing which branch of the network to try and attach a block to. In a short time, one of the two branches prevails over the other and the blockchain resolves the fork. The value of $j$ must therefore be optimised so as to reward as many miners as possible, without disproportionately increasing the number of communications or forks.

Note that to build our new system we followed Shi's smart idea, but making an essential change. Shi proposed to insert the new block once a single miner has reached a fixed number of solutions to the problem. 

By doing so, however, large miners would have an unfair advantage and would tend to have a reward greater than their weight.

Our solution of inserting the block when the entire network finds a fixed number of solutions to the problem, in our case a fixed number $j$ of nonces, and not when a single miner finds $j$ nonces, allows us to avoid this problem and share the reward fairly.

At the end we observe that our new protocol is no more sensitive than the standard Bitcoin proof-of-work protocol to selfish-mining attacks \cite{Eyal-Sirer2}. 

\section{Analysis of the new systems}

To model the Bitcoin protocol, we can use a Bernoulli process $\{X_n\}_{n \in \mathbb{N}}$, where $X_n$ are independent and identically distributed Bernoulli random variables of parameter $p = \frac{1}{600 H}$ and $H$ is the total hashrate of Bitcoin. Consider now the random variable $S$, that models the time of first success of a Bernoulli process:
\[S=\min\{n \in \mathbb{N}: X_n=1\}.\]
Clearly, $S$ is a geometric random variable of parameter $p$ and its expected value is  $\mathbb{E}(S)= \frac{1}{p} = 600\; H$, the average number of hashes required to insert a new block into the Bitcoin blockchain. Since the network processes about H hashes in one second, this average waiting time is about 10 minutes. The variance of $S$ is also straightforward and it is equal to
$$\mathbb{V}(S)=\frac{1 - p}{p^2} \sim \frac{1}{p^2},$$
so the standard deviation of $S$ is of the same order than its expected value.

Similarly, to model our new protocol, we can use a Bernoulli process $\{Y_n\}_{n \in \mathbb{N}}$, where $Y_n$ are independent and identically distributed Bernoulli random variables of parameter $jp$ and $j < 600 \; H$. Let $T_j$ be the random variable that models the time of $j$-th success of the Bernoulli process $\{Y_n\}_{n \in \mathbb{N}}$:
$$T_j=\min \{ n \in \mathbb{N}: Y_1+ \ldots +Y_n =j\}.$$

Notice that $T_j$ is distributed as a negative binomial random variable with parameters $(j,jp)$, where the first parameter models the required number of successes, while the second parameter models the success probability in each Bernoulli trial. We know a closed formula to compute its expected value and its variance:
\[ \mathbb{E}(T_j)=j \; \frac{1}{jp} = 600\; H, \hspace{1cm} \mathbb{V}(T_j)=j \;\frac{1-jp}{(jp)^2}=  \frac{600^2\; H^2}{j} - 600\; H.\]

We thus have that the new protocol has an average waiting time for inserting a block identical to that of the current Bitcoin protocol, while the variance of this waiting time is smaller for each $j\geq 2$ and decreases as $j$ increases. 

To be more explicit, we can compute the probability of inserting a block between 9 and 11 minutes (540\;H and 660\;H hashes, respectively) in the actual Bitcoin protocol and compare it with the same probability of our models with some fixed $j$ values, like $j= 10$ or $j= 100$: 
\[\mathbb{P} \left[540\;H < S < 660\;H\right] \sim 7.31\%, \]
\[\mathbb{P} \left[540\;H < T_{10} < 660\;H\right] \sim 24.69\%, \]
\[\mathbb{P} \left[540\;H < T_{100} < 660\;H\right] \sim 72.36\%.\]

Switching to the generalised protocol would therefore make it rarer for a block to be inserted in a much smaller or much larger time than the expected average time of 10 minutes. This considerable reduction in the variance of the block insertion time would significantly improve the security of the system, since it is well known that the high variance of block insertion time is at the root of critical attacks on PoW blockchains \cite{Bobtail}.

Let us now calculate how the profit is distributed among the different miners in the new protocol compared to the current Bitcoin protocol.

Let $q$ be the probability of a generic Bitcoin miner to find a suitable nonce such that the hash of the block header is less than the network target. Note that this probability does not depend on the target at all, but only on the miner's weight (i.e., the ratio of his hashrate to the network's hashrate). A reasonable estimate of this probability can be calculated from past block insertion statistics. Let $M = 52560$ be the average number of blocks mined in a year. Define now the Bernoulli process $\{B_k\}_{k \in \mathbb{N}}$, where $B_k$ are independent and identically distributed Bernoulli random variables of parameter $q$, that model the probability of adding a block for the considered miner. 

Let
\[U= 6.25 \sum_{k=1}^M B_k\]
be the profit of the miner in a year. 

It is clear that
\[\mathbb{E}(B_k)= q, \ \  \mathbb{V}(B_k)= q (1-q)\]
\[\mathbb{E}(U)= 6.25\; M\; q, \ \  \mathbb{V}(U)= (6.25)^2 \; M \, q(1-q).\]

On the other side, to model our protocol, we define a Bernoulli process $\{C_i\}_{i \in \mathbb{N}}$,  where $C_i$ are independent and identically distributed Bernoulli random variables of parameter $q$, and a Binomial random variable $N_k$ 
\[N_k= \sum_{i=1}^j C_i,\]
that counts the number of nonces found for the block $k$ by the considered miner. Finally, let $V$ be the random variable defined as
\[V= 6.25 \sum_{k=1}^M  \frac{N_k}{j}, \]
that models the profit of the miner in a year. 

Thus, we have
$$\mathbb{E}(N_k) = j q, \hskip1cm \mathbb{V}(N_k) = j q(1-q)$$
$$\mathbb{E}(V) = 6.25\; M\; q, \hskip1cm \mathbb{V}(V) = (6.25)^2\; M \; \frac{q(1-q)}{j}. $$

We have obtained that, in the new protocol, the profit of each miner has the same expected value as in the current Bitcoin protocol and the variance of this profit is always lower and decreases as $j$ increases.

To evaluate in a further way the improvement that would be made with the new system, it may be convenient to estimate the probability for a miner of weight $p$ to obtain an annual profit not too much lower than the expected value. For example, in Bitcoin, a miner has a greater than 95\% chance of making an annual profit $U$ greater than 70\% of the expected profit if it holds a weight equal to $0.0005304$, that is
\[\mathbb{P}[ U > 0.7 \ \mathbb{E}(U)]\geq0.95 \qquad \qquad \text{if } p = 0.0005304.\]

The same estimation can be repeated for the new model, for example, if $j=10$, to get the same enterprise risk it is sufficient for a miner to have weight $p=0.00005305$, i.e.
\[\mathbb{P}[ V > 0.7 \ \mathbb{E}(V)]\geq0.95 \qquad \qquad \text{if } p = 0.00005305.\]

More generally, the weight a miner has on Bitcoin can be divided by $j$ on the new protocol, so as to maintain essentially the same probability of making a suitably high profit.

Thus, in the new system, one could create a mining company with the same business risk but with a lower initial investment by one order of magnitude, in the case $j=10$, or by two orders of magnitude, in the case $j=100$. 

\section{Conclusions}

The new proof-of-work system we presented would regularize the variance of block entry, and as we have seen this fact would improve the security of the system. A second result is the notable advantage of distributing the rewards more widely among the miners, which would greatly reduce the variance of profit distribution.

The essential characteristics of Bitcoin are all maintained and this allows to avoid new ways of attacking the system (we have the same attacks as in the Bobtail model \cite{Bobtail}). The adoption of this new protocol would therefore allow for a significant reduction in the business risk associated with mining and would allow for the start of this activity with a smaller investment in computing resources than the current Bitcoin, without significantly reducing the security of the system.

The level of business risk reduction is proportional to the parameter $j$, the number of nonces that the network must find for a block insertion. The larger $j$ is, the smaller the variance of the profit is.

The negative consequence is that increasing $j$ also increases the probability that more miners arrive almost simultaneously to propose a new block to be inserted in the blockchain. The ingenious system devised by Satoshi Nakamoto \cite{Nakamoto} to manage these critical situations would surely work also in our new system, but it is not entirely clear up to which values of $j$ the network would be able to manage them efficiently and above which values an excessive number of forks could be created and make the system unstable.

In a forthcoming work we intend to study the problem of choosing the parameter $j$ and also evaluate a dynamic optimization 
in a similar way to how Bitcoin manages the network target.

\end{document}